\documentclass[11pt]{article}
\usepackage{cite}
\usepackage[useregional]{datetime2}
\usepackage{babel}
\usepackage{graphicx}
\usepackage{epigraph}

\usepackage[hidelinks]{hyperref}
\usepackage{textcomp}

\usepackage{tikz}
\usepackage{varwidth}

\usepackage{float}
\restylefloat{table}

\usepackage{multirow}
\usepackage{geometry}
\usepackage{setspace}
\usepackage{microtype}

\newcommand{\etal}{\textit{et al. }}
\newcommand{\tild}{\raise.30ex\hbox{$\scriptstyle\sim$}}

\begin{document}

\title{Understanding and Avoiding AI Failures: \\A Practical Guide}
\author{Heather M. Williams \\ University of Louisville \and Roman V. Yampolskiy \\ University of Louisville}
\date{\today}

\maketitle

\begin{abstract}
As AI technologies increase in capability and ubiquity, AI accidents are becoming more common.  Based on normal accident theory, high reliability theory, and open systems theory, we create a framework for understanding the risks associated with AI applications.  In addition, we also use AI safety principles to quantify the unique risks of increased intelligence and human-like qualities in AI.  Together, these two fields give a more complete picture of the risks of contemporary AI.  By focusing on system properties near accidents instead of seeking a root cause of accidents, we identify where attention should be paid to safety for current generation AI systems.
\end{abstract}


\section{Introduction}
\label{sec:introduction}

With current AI technologies, harm done by AIs is limited to power that we put directly in their control. As said in
~\cite{yam2018historic}, ``For Narrow AIs, safety failures are at the same level of importance as in general
cybersecurity, but for AGI it is fundamentally different.'' Despite AGI (artificial general intelligence) still being
well out of reach, the nature of AI catastrophes has already changed in the past two decades. Automated systems are now
not only malfunctioning in isolation, they are interacting with humans and with each other in real time. This shift has
made traditional systems analysis more difficult, as AI has more complexity and autonomy than software has before.

In response to this, we analyze how risks associated with complex control systems have been managed historically and
the patterns in contemporary AI failures to what kinds of risks are created from the operation of any AI system. We
present a framework for analyzing AI systems before they fail to understand how they change the risk landscape of the
systems they are embedded in, based on conventional system analysis and open systems theory as well as
AI safety principles.

Finally, we present suggested measures that should be taken based on an AI system's properties. Several case studies
from different domains are given as examples of how to use the framework and interpret its results.

\section{Related Work}

\subsection{Early History}

As computer control systems increased in complexity in the 70's and 80's, unexpected and sometimes catastrophic
behaviour would emerge from previously stable systems \cite{anderson2005control}. While linear control systems (for
example, a thermostat) had been used for some time without unexpected behaviour, adaptive control systems created novel
and unexpected problems, such as ``bursting''. As described in~\cite{anderson2005control}, bursting is the phenomenon
where a stable controller would function as expected for a long time before bursting into oscillation, then returning to
a stable state. This is caused by the adaptive controller not having a rich enough input during the stable period to
determine the unknown coefficients of its model correctly, causing the coefficients to drift. Once the system enters
oscillation, the signal again becomes rich enough for the controller to correctly estimate the unknown coefficients and
the system becomes stable again. The increased complexity of the more advanced technology (dynamic controller instead of
a static controller) introduced a dynamic not present in previous technologies, and incomprehensible to an operator not
familiar with this behavior. Worse, since this behavior only happens when the controller is controlling the real world
plant, designers had no way of predicting this failure mode. Bursting can be reduced using specifically engineered
safety measures or more complex controllers (which bring even more confounding problems), but still demonstrates that
increases in complexity tends to increase risk.

\subsection{Normal Accident Theory}

\renewcommand{\epigraphflush}{center}
\renewcommand{\textflush}{flushepinormal}
\setlength \epigraphwidth {5in}
\epigraph{ One of the principal values of `normal accident' analysis and case descriptions
is that it helps to develop convincing materials to counter the naive, perhaps wistful or short-sighted, views of
decision-makers who, due to institutional pressure, desperation or arrogance, are temped to make unrealistic
assumptions about the systems they direct but for which they have only nominal operational responsibility. }
{\textit{Todd R. La Porte}~\cite{laporte1994strawman}}

Risk of failure is a property inherent to complex systems, and complex systems are inherently
hazardous~\cite{cook1998complex}. At a large enough scale, any system will produce ``normal accidents''. These are
unavoidable accidents caused by a combination of complexity, coupling between components, and potential harm. A normal
accident is different from more common component failure accidents in that the events and interactions leading to
normal accident are not comprehensible to the operators of the system~\cite{perrow1984living}. Increasing the complexity
and broadening the role of AI components in a system decreases comprehensibility of the system, leading to an increase
in normal accidents.

In 1984, Charles Perrow published ``Normal Accidents''~\cite{perrow1984living} which laid the groundwork for NAT (normal
accident theory). Under NAT, any system that is tightly coupled and complexly interactive will inevitably experience a
system accident. Decentralization reduces coupling and increases complexity, while centralization decreases complexity
but also increases coupling. Thus, since an organization cannot be both centralized and decentralized, large
organizations will harbor system properties that make them prone to normal accidents.

\subsection{High Reliability Theory}

High reliability theory was developed to explain the incredible ability of certain organizations to function without
major accidents for long periods of time. In~\cite{weick1999reliability}, Weick identifies several common traits shared
by high reliability organizations (HROs): a strategic prioritization of safety, careful attention to design and
procedures, redundancy, decentralized decision making, continuous training
often through simulation, strong cultures that encourage vigilance and responsiveness to potential accidents, and
a limited degree of trial-and-error learning.

High reliability organizations manage the apparent paradox proposed in normal accident theory by having traits of both
centralization and decentralization. Decision making is decentralized, allowing for decoupling, while policy and
cultural factors are highly centralized, allowing for the unification of goals and attention to safety. This ability to
be simultaneously centralized and decentralized through common culture and goals is present not only in HROs but in
collectivist cultures, demonstrated by the tendency of members of these cultures to cooperate in social
dilemmas~\cite{parks1994collectivist}~\cite{hofstadter1983lottery}.

\subsection{NAT-HRT Reconciliation}

Normal accident theory holds that for industries with certain system properties, system accidents are inevitable.
Meanwhile, high reliability theory makes the observation that there are many exceptions to this, and there are common
traits shared by these HROs that can be studied and understood as indicators of
reliability. In~\cite{shrivastava2009normal}, Shrivastava \etal analyze how the two theories appear to be in conflict
then reconcile them by looking at how debates between the two sides neglect the importance of time in understanding
accidents.

Normal accident theory states that a system has to choose to trade off between centralization (which allows for
organization-aligned action and less chaos) and decentralization (which enables operators to quickly make decisions).
High reliability theory considers that it is possible to escape this apparent paradox by allowing operators a high level
of autonomy (decentralized decision making) while also putting a focus on cultural aspects that improve safety and
reliability (centralized goals).

Shrivastava \etal use the Swiss cheese model (SCM) to explain the importance of time in the occurrence of accidents,
even in systems that are stable over time. In the SCM, the layers of safety are modeled as slices of cheese with holes
in them representing weak points in each layer of safety. Over time, holes change shape and move around. Eventually, no
matter how many slices of cheese (layers of safety) there are, the holes will align allowing a straight shot through all
of the slices of cheese (an accident occurs)~\cite{nushi2018pandora}. However, the SCM model only demonstrates that
accidents are inevitable, and ``inevitability is immaterial for practical purposes'' as Shrivastava \etal state, since
the time scales involved for some systems may be far longer than the system is expected to operate.

Disaster incubation theory (DIT) is introduced as the final piece needed to reconcile normal accident theory and
high reliability theory. DIT describes how organizations gradually migrate to the boundary of acceptable behavior, as
good safety records drive up complacency and deviance is
normalized~\cite{vaughan2005slippery}~\cite{rasmussen1997modelling}. DIT was considered only useful in hindsight by
proponents of NAT, so Snook~\cite{snook200friendly} investigated accidents with this in mind and created the framing of
``practical drift''. 
This is the ``slow steady uncoupling of local practice from written
procedure''~\cite{snook200friendly} which leads an
initially highly coupled system to become uncoupled as operators and managers optimize their processes to be more
efficient, deviating from procedure. Then, if the system is required to become tightly coupled again, the operators are
ill-prepared for this increase in coupling and a system accident can occur.

Through the lens of disaster incubation theory and practical drift, Shrivastava \etal explain how NAT and HRT work to
compliment each other to explain how accidents take place and are avoided. The time period being considered by HRT takes
place while the system is still a high reliability organization. The culture and procedures put in place are working
correctly, coupling is high, and complexity is manageable to the well trained operators. Over time, however, practical
drift decouples the system and reliability decreases. If the organization is a high reliability organization,
degradation is limited and incidents can still be managed. Accidents that take place during this period of time are
within the scope of HRT. However, there is the possibility for an unlikely event to lead from this steady decline of
reliability to a normal accident. If the system suddenly has to become more coupled (for instance, during a special
mission or to react to an incident breaking down multiple layers of safety), it is ill prepared to do so. At
this point, NAT's trade-off between coupling and complexity becomes important, and the perceived complexity of the
system increases drastically, making safe operation impossible, leading to a normal accident. This can only happen after
a great decrease in coupling from the initial (designed) state of the system, so proponents of HRT would say that the
accident was only able to take place due to the system no longer acting as a high reliability organization.

High reliability theory explains how organizations resist practical drift, and the accidents that
happen when practical drift leads to a breakdown of high reliability practices. Normal accident theory is useful when
practical drift has lead to a great degree of decoupling, then a sudden change in situation (which may be an intentional
operation or an unexpected incident) requires increased coupling, which the system is (surprisingly, to operators) no
longer able to handle without increasing complexity beyond manageable levels.

\subsection{Lethal Autonomous Weaponry}

The introduction of lethal autonomous weaponry~\cite{carvin2017normal} increases the danger of normal accidents not
because it provides new kinds of failure or novel technologies but because of the drastically increased potential
harm. A machine which kills when functioning correctly is much more dangerous in an accident than one which only does
harm when malfunctioning. By increasing the level of complexity and autonomy of weapons systems, normal accidents
involving powerful weapons becomes a possibility.

\subsection{Robustness Issues for Learned Agents}

In~\cite{uesato2018adversarial}, Uesato \etal train a reinforcement learner in an environment
with adversarial perturbations instead of random perturbations. Using adversarial perturbations, failure modes that
would be extremely unlikely to be encountered otherwise were detected and integrated into training. This shows that AI
trained to be ``robust'' by training in a noisy environment may still have catastrophic failure modes that are not
observed during training, which can spontaneously occur after deployment in the real world. Adversarial training is a
tool to uncover and improve these issues. However, it is only an engineered safety measure over the deeper issue of
black box AI, which are not characterized of their entire input space.

Image classifiers famously fail when faced with images that have been modified by as little as one
pixel~\cite{su2019onepix}, called adversarial examples. Despite performing well on training data, test data,
and even data from other datasets, image classifiers can be made to reliably misclassify images by changing the images
so slightly that the alterations are invisible to the casual observer. In~\cite{ilyas2019bugs}, Ilyas \etal argue that
these misclassifications are not due to a simple vulnerability, but due to image classifiers' reliance on
\textit{non-robust features}. These are features which are not apparent to the human eye but can be used to accurately
classify images, even those outside of the original dataset. Non-robust features are transferable to other datasets and
the real world. However, they are also invisible to the human eye and can be altered without noticeably changing the
appearance of the image. Classifiers with only robust features can be created through \textit{robust
training}~\cite{madry2019resistant}, but they suffer from decreased accuracy. Thus, non-robust features are a useful
mechanism to achieve high accuracy, at the cost of vulnerability to adversarial attacks.

The difference between robust and non-robust features is strictly human-centric. Ilyas \etal frame this as an alignment
problem. While humans and image classifiers are superficially performing the same task, the image classifiers are doing
it in a way the is incomprehensible to humans, and can fail in unexpected ways. The misalignment between
the objective learned from the dataset and the human notion of an image belong to any particular class is the underlying
cause for the effectiveness of adversarial examples.

This is a useful framing for other domains as well. A reinforcement learner achieving impossibly high scores by hacking
its environment doesn't ``know'' that it's breaking the rules --- it is simply doing what was specified, and incredibly
well. Just as a robust classifier loses some accuracy from being disallowed non-robust features, a reinforcement learner
that is prevented from reward hacking will always obtain a lower reward. This is because while the
designer's goal is to create a useful agent, the agent's goal is to maximize reward. These two are always
misaligned, a problem referred to as the alignment problem \cite{taylor2020alignment}.

\subsection{Examples of AI Failures}
\label{sec:examples}

Large collections of AI failures and systems to categorize them have been created
before~\cite{yam2018historic}~\cite{scott2020classification}. In~\cite{scott2020classification}, the classification
schema details failures by problem source (such as design flaws, misuse, equipment malfunction, etc.), consequences
(physical, mental, emotional, financial, social, or cultural), scale of consequences (individual, corporation, or
community), and agency (accidental, negligent, innocuous, or malicious). It also includes preventability and software
development life-cycle stage.

The AI Failures Incident Database provides a publicly accessible view of AI failures~\cite{incidentwebsite}. There are
92 unique incidents the have been reported into it. A review of the agency (cause) of each one is not listed on the
website, so we give an overview here. Of the 92 incidents, 8 have some degree of malicious intent. Three are cases
where social media users or creators manipulated AI to show inappropriate content or cause bots to produce hate speech.
Two are incidents of hacking: spoofing biometrics and stealing Etherium cryptocurrency. One is the use of AI
generated video and audio to misrepresent a public official. The last two are the only cases where AI could be seen as a
malicious agent. In one case, video game AI exploited a bug in the game to overpower human players. In another, bots
created to edit Wikipedia competed in a proxy war making competing edits, expressing the competing desires of their human
creators. If these examples are representative, then a majority of AI incidents happen by accident, while less than 10\%
are the result of malicious intent. The 2 examples of AI malicious intent intent can be ascribed to AI given goals which
put them in opposition of others: countering edits in one case, and waging warfare (within a video game) in the other.

\subsection{Societal Impact of AI}

The proliferation of AI technologies has impacts in our socioeconomic systems and environment in complicated ways, both
positive and negative~\cite{jayden2018sustainability}. AI has the ability to make life better for everyone but also to
negatively impact many by displacing workers and increasing wealth disparity~\cite{hagerty2019social}. This is just one
example of AI interacting with a complicated system (in this case, the job market) to have large scale consequences.
While the effects of AI technologies on society and culture are outside of the scope of this paper, we expect that AI
will continue to increase in ubiquity and with it the increased chance for large scale AI accidents.

\subsection{Engineered Interpretability Measures}

In~\cite{nushi2018pandora}, Nushi \etal present Pandora, a state of the art image captioning system with novel
interpretability features. It clusters its inputs, and predicts modes of failure based on
latent features. The image captioning system is broken down into three parts, the object detector, the language model,
and the caption reranker. Pandora is able to predict error types based on the inputs to each of these models as well as
the interactions between small errors in each part of the system accumulating to a failure. This feature is of
particular interest, as it is a step towards improved interpretability for complex systems. Pandora is remarkable for
creating explanations for a black-box model (an artificial neural network) by training a trivially interpretable model
(a decision tree) on the same inputs with failure modes as its output.

In~\cite{das2020explainable}, Das \etal present an explanation generating module for a simulated household robot. A
sequence-to-sequence model translates the robot's failure state to a natural language explanation of the failure and why
it happened. Their results are promising --- the explanations are generated reliably (\tild90\% accuracy) and the model only
makes mistakes within closely related categories. The generated explanations improve the accuracy of an inexperienced
user for correcting the failure, especially when the explanation contains the context of the failure. For example,
instead of the robot stating only ``Could not move its arm to the desired object.'', it also gives a reason: ``Could not
move its arm to the desired object because the desired object is too far away.''

Both of these papers offer major contributions to making AI components and systems comprised of them safer and more
understandable. However, measurement mechanisms can create even greater degrees of confusion when they fail, and can
delay or confound the diagnosis of a situation. When two pressure gauges were giving conflicting information at the
Three Mile Island accident~\cite[p. 25]{perrow1984living}, they increased confusion instead of providing information.
An AI which says nothing except that a malfunction has occurred is easier to fix than one that gives misleading
information. If the accuracy is high enough, if the explanations are bounded for how incorrect they can be, or if the
operator knows to second-guess the AI, then the risks created by an explanation system are limited. The utility of
providing explanations of failures, as demonstrated in~\cite{das2020explainable}, is a compelling reason to add
interpretability components despite the additional complexity and unique risks they create. Further study into real
world applications of these systems is needed to understand the pros and cons of implementing them.

\subsection{Faulty Post-hoc Explanations in Humans}

Experiments done on split-brain patients~\cite{gazzaniga1998split} show that the left hemisphere of the brain will
generate incorrect explanations for the right hemisphere's actions with high confidence when the same information is not
available to both hemispheres. This indicates that explainability is not a function of general intelligence but instead
an additional module the has evolved in the human brain, likely as a mechanism to facilitate social
cooperation~\cite{haidt2012righteous}. Thus we should be skeptical of plausible sounding explanations generated by AI,
as humans provide a model for generating plausible but incorrect explanations, which obscure the hidden information that
explanations are meant to express.

\subsection{The AI Accident Network}

Attributing fault is difficult when AI does something illegal or harmful. Punishing the AI or putting it in jail
would be meaningless, as our current level of AI lacks personhood. Even if trying an AI for crimes was deemed
meaningful, traditional punishment mechanisms could fail in complex ways~\cite{bostrom2014handbook}.
Instead, the blame must fall on some human or corporate actor. This is difficult because the number of parties
responsible for the eventual deployment of the AI into the world could be large: the owner of the AI, the operator, the
manufacturer, the designer, and so on. Lior \cite{lior2020network} approaches this problem from a variety of legal
perspectives and uses a network model of all involved agents to find the party liable for damages caused by an AI.
A compelling anecdote introduces the problem: a child is injured by a security robot at a mall. Who is liable
for the child's injuries? Lior frames liability from the paradigm of ``nonreciprocal risk'' which states that ``[if] the
defendant has generated a `disproportionate, excessive risk of harm, relative to the victim's risk-creating activity',
she will be found liable under this approach.''

Lior argues against other definitions of liability and that nonreciprocal risk is the best way to determine liability in
the context of AI accidents. Lior then describes how the importance of different agents in an accident can be understood
using network theory. By arranging the victims, AI, and responsible or related parties into a network, network
theory heuristics can be used to locate parties liable for damages. These heuristics include the degree of a node (its
number of connections), measures of centrality of a node in the network, and others.

As AI technologies are proliferated, AI accidents are happening not only in commercial settings (where damages are
internal to the corporation and assigning blame is an internal matter) but in public settings as well. Robots causing
injuries to visitors and autonomous vehicles being involved in crashes are both novel examples of this. For companies
planning on using AI in externally deployed products and services, understanding how AI liability is going to be treated 
legally is crucial to properly managing the financial and ethical risks of deployment.

\subsection{AI Safety}

The field of AI safety focuses on long-term risks associated with the creation of AGI (artificial general intelligence).
The landmark paper ``Concrete Problems in AI Safety''~\cite{amodei2016concrete} identifies the following problems for
current and future AI: avoiding negative side effects, avoiding reward hacking, scalable oversight, safe exploration,
and robustness to distributional change. Other topics identified as requiring research priority by other authors include
containment \cite{babcock2016containment}, reliability, error tolerance, value specification
\cite{soares2017foundations}~\cite{hadfieldmenell2020inverse}, and
superintelligence~\cite{bostrom2001extinction}~\cite{bostrum2014superintelligence}. These topics are all closely related
and could all be considered an instance of the ``Do What I Mean'' directive~\cite{yudkowsky2011dwim}. We will explore
the five topics from~\cite{amodei2016concrete} with examples of failures and preventative measures where applicable.

\subsubsection{Avoiding Negative Side Effects}

This problem has to do with things that are done by accident or indifference by the AI. A cleaning robot knocking
over a vase is one example of this. Complex environments have so many kinds of ``vases'' that we are unlikely to be able to
program in a penalty for all side effects~\cite{amodei2016concrete}. A suite of simulated environments for testing AI
safety, the AI Safety Gridworlds, includes a task of moving from one location to another without putting the environment
in an irreversible state~\cite{leike2017gridworlds}. Safe agents which avoid side effects should prefer to avoid this
irreversible state. To be able to avoid negative side effects, an agent has to understand the value of everything in its
environment in relation to the importance of its objective, even things that the reward function is implicitly
indifferent towards. Knocking over a vase is acceptable when trying to save someone's life, for example, but knocking
over an inhabited building is not. Many ethical dilemmas encountered by people are concerned with weighing the
importance of various side effects, such as environmental pollution from industrial activity and animal suffering from
farming. This is a non-trivial problem even for humans~\cite{irving2019social}. However, using ``common
sense'' to avoid damaging the environment while carrying out ordinary tasks is a realistic and practical goal with our
current AI technologies. Inverse reward design, which treats the given reward function as incomplete, is able to avoid
negative side effects to some degree~\cite{hadfieldmenell2020inverse}, showing that making practical progress in this
direction is achievable.

\subsubsection{Avoiding Reward Hacking}
\label{sec:rewardhacking}

Most AI systems designed today contain some form of reward function to be optimized. Unless designed with safety
measures to prevent reward hacking, the AI can find ways to increase the reward signal without completing the objective.
These might be benign, such as using a bug in the program to set the reward to an otherwise unattainably high
value~\cite{chrabaszcz2018qbert}, complicated such as learning to fall over instead of learning to
walk~\cite{lehman2018surprising}, or dangerous, such as coercing human engineers to increase its reward signal by
threatening violence or mindcrime \cite{bostrum2014superintelligence}.

Agents that wish to hack their rewards can do so by breaking out of their containers, whether they are within simple
training environments or carefully engineered prisons with defenses in depth~\cite{babcock2016containment}. When
researching a novel AI technology that has a risk of creating AGI, researchers must use safety measures to prevent
potential AGI from escaping confinement. This is not common practice as AI today are not believed to be capable of
escaping the simulations they are tested in. However, humans and AI have both broken confinement in games that are used
to benchmark AI. In 2018, an AI designed to maximize the score in a video game instead performed a series of seemingly
random actions which caused the score to increase rapidly without continuing to progress in the
game~\cite{chrabaszcz2018qbert}. In this example, the AI is supposed to be using the game controller output to move the
character around the screen to play the game and maximize the score, but it instead learned to hack the game and obtain
a much higher score than is conventionally possible. Through an exploit in the game's programming the AI managed to
directly modify its score instead of taking actions within the simulation as expected. In 2014, an exploit in Super
Mario World was used to modify memory and jump directly to the end screen of the game a few minutes into play. This also
demonstrates an agent (in this case, a computer assisted human) breaking a level of containment through a sophisticated
attack.

In light of these examples, it is possible to gain arbitrary code execution abilities from within a sandboxed game
environment and these kinds of exploits may be present inside any program which is meant to limit the output space of
an AI. Virtual machines, often used as an additional layer of security, are also susceptible to a wide range of
exploits~\cite{reuben2007virtual}. This illustrates a more general concern which is the AI acting outside of the output
space that it was designed to work with, seen in many of the AI failures in~\cite{yam2018historic}.

Adversarial reward functions, where the reward function is learned and has a competing objective with the AI, can
decrease reward hacking~\cite{fu2018adversarial}. Another mitigation strategy is the use of tripwires. For instance, if
the AI gets a much higher score than expected, it can be assumed that it has broken containment and is modifying the
reward directly. This can be used as a tripwire which, when triggered, deactivates the AI.

\subsubsection{Scalable Oversight}

Designers only have limited resources to monitor the actions of the AI. External factors are inevitably forgotten about
(worsening side effects) and reward functions fail to perfectly capture what we really want (worsening reward hacking).
We cannot monitor every action taken by an AI at every moment because they operate too fast or in a way that's too
complex for us to understand, and because the goal is often to automate a task specifically so it can be completed
without constant human oversight. Semi-supervised learning~\cite{amodei2016concrete} is a first step towards scalable
oversight as it allows labeled and unlabeled data to be used to train an AI. In an online learning context, this means
that the AI can learn by doing the task while only occasionally needing feedback from a human expert. Semi-supervised
learning is useful in terms of data efficiency and is a promising avenue for creating scalable oversight for AGI.

When dealing with AGI, scalable oversight is no longer an issue of data efficiency or human effort but of safety. At
some point, the AGI will be intelligent enough that we are not able to tell if it is acting in our best interest or
deceiving us while actually doing something unsafe. Scalable oversight is required to make AGI safe. One phrasing of a
scalable oversight from the perspective of the AGI is ``If the designers understood what I was doing, they would approve
of it.''

\subsubsection{Safe Exploration}

Reinforcement learning algorithms, such as epsilon-greedy and Q-learning, occasionally take a random action as a way to
explore their environments. This is effective, especially with a small environment and a very large amount of training
time, but it is not safe for real world environments. You don't want your robot to try driving over a cliff for the sake
of exploration. Curiosity-driven exploration is able to efficiently explore video game environments by using an
intrinsic reward function based on novelty~\cite{pathak2017curiosity}. This provides great benefits to the learner, but
can't ensure safety. For example, the agent only learns to avoid death because restarting the level is ``boring'' to it.
A backup policy which can take over when the agent is outside of safe operating conditions can allow for safe and
bounded exploration, such as an AI controlled helicopter which is switched to a hover policy when it gets too close to
the ground or travels too fast \cite{garcia2014explore}.

Robotics controlled by current narrow AI need to be designed to avoid damaging expensive hardware or hurting people.
Safe exploration becomes more difficult with increased intelligence. A superintelligent AGI might kill or torture humans
to understand us better or destroy the Earth simply because such a thing has never been done before.

\subsubsection{Robustness to Distributional Change}

AI that has been trained in one environment or dataset may fail when its use is greatly different from its training.
This is the cause AI obtaining good accuracy on curated datasets but failing when put into the real world.
Distributional shift can also be seen in the form of racial bias of some AI \cite{hamilton2020obama}.

\subsection{The Interpretation Problem}

The difficulty of converting what we want into rules that can be carried out by an AI has been named the
``interpretation problem''~\cite{badea2021interpretation}. This problem arises in domains outside of AI, such as sports
and tax law. For sports, the rules of a game are written to reflect a best effort representation of the spirit of the
game to bring out creativity and skill in the players. The players, however, seek only to win and will sometimes create
legal tactics which ``break'' the game. At this point, to keep the game from stagnating, the rules are modified to
prohibit the novel game-breaking strategy. A similar situation takes place in tax law, with friction between lawmakers and
taxpayers creating increasingly complex laws.

In AI safety, the designers knows what is and isn't morally acceptable from their point of view. They then design rules
for the AI to abide by while seeking its objective and craft the objective to be morally acceptable. The AI, however,
only has access to the rules and not to the human values that created them. Because of this, it will misinterpret the
rules as it pursues its objective, causing undesirable behaviour (such as the destruction of humanity). Humans do not
learn morality from a list of rules, but from some combination of innate knowledge and lived experience. Badea \etal
suggest that the same applies to AI. Any attempt to create a set of rules to constrain the AI to moral behaviour will
fail due to the interpretation problem. Instead, we need to figure out how we can show the AI human moral values
indirectly, instead of having to state them explicitly. Since our goals and ``common sense'' as applied to any real
world environment are too complex to write down without falling prey to the interpretation problem, an alternative
approach where the AI is able to acquire values from us indirectly is required instead. Badea \etal do not provide
details on how this could be accomplished, but there are some technologies which are very similar in nature to the
proposed value learning system. Inverse reinforcement learning~\cite{menell2016inverse} is able to learn objectives
from human demonstrations, and can learn goals that would be difficult to explicitly state. Future developments in this
area will be needed to create moral AI.

\section{Classification Schema for AI Systems}
\label{sec:classification}

To better understand an AI system and the risks that it creates, we have created a classification schema for AI systems.
We present a tag based schema similar to the one presented in~\cite{scott2020classification}. Instead of focusing on AI
failures in isolation, this schema includes both the AI and the system the AI is embedded in, allowing for detailed risk
analysis prior to failure. We pay particular attention to the orientation of the AI as both a system with its own
inherent dangers and as a component in a larger system which depends on the AI. Any analysis attempting to divide a
system into components must acknowledge the ``ambiguities, since once could argue interminably over the dividing line
between part, unit, and subsystem''~\cite{perrow1984living}. The particular choice of where to draw the line between the
AI, the system, and the environment are important to making a meaningful analysis, and requires some intuition. Because
of this, our classification system encourages repeating parts of it for different choices of dividing lines between the
AI and the system to encourage finding the most relevant framing.

To characterize the risk of an AI system, one factor is the dangers expressed by the larger system in which the AI is a
component. In an experimental setting, a genetic algorithm hacking the simulator can be an amusing
bug~\cite{lehman2018surprising}, but a similar bug making its way into an autonomous vehicle or industrial control
system would be dangerous. Understanding the risks involved in operating the system that the AI belongs to is critical
in understanding the risk from an AI application.

The output of an AI must be connected to some means of control to be useful within the system. This can take many forms:
indirectly, with AI informing humans who then make decisions; directly with an AI controlling the actuators of a
robot or chemical plant; or though information systems, such as a social media bot that responds to users in real time.
Any production AI system has some degree of control over the world, and it may not be clear where the effects of the AI
take place. The connection between the output of a component (the AI) and other components is an instance of coupling.
Because there are multiple components affected by the AI and those components are themselves coupled with still other
components, we frame this problem in terms of the AI, which has its own properties, and a series of targets which the
AI can affect. For example, the output from an AI in a chemical plant can be framed in numerous ways: the software
signal with the computer the AI is running in, the electrical output from the controller, the actuation of valve, the
rate of fluid flowing through the valve, or a variable in the chemical reaction taking place downstream. Identifying
the target of the AI requires context, and all of these targets have unique consequences the might be overlooked in
analyzing just one of them. The classification schema allows for as many targets as needed when analyzing a system.

Observability of the warning signs of an accident is important as it allows for interventions by human operators, a
crucial tool in preventing accidents (``People continuously create safety''~\cite{cook1998complex}). The likelihood of
timely human interventions depends on four things:

\begin{itemize}
    \item Time delays between AI outputs and the effects of the target.
    \item Observability of the system state as it is relevant to potential accidents.
    \item Frequency and depth of attention paid by human operators.
    \item Ability of operators to correct the problem once it has been identified.
\end{itemize}

The time delay between an AI creating an output and that output affecting the target is essential to preventing
accidents. Tightly coupled systems with short time delays (such as automated stock trading) are more hazardous because
the system can go from apparent normalcy to catastrophe faster than operators can realize that there is something
wrong~\cite{andrei2017flash}. Observability and attention from human operators are needed for these time delays to be
an effective component of safety. As the level of automation of a system increases, human operators become less
attentive and their understanding of its behavior decreases~\cite{bainbridge1983ironies}. Reliance on automated systems
decreases an operator's ability to regain control over a system if an accident requires manual control. For example, if
an autonomous driving system fails, the driver, now less familiar with driving, has to suddenly be in manual control.
Together, observability, human attention, and human ability to correct possible failures in the system all make up a
major factor in whether or not a malfunction leads to an accident.

For a given choice of target being controlled by the AI, there is maximum conceivable amount of damage that can be done
by malicious use of that target. We use the figure as a cap as to the amount of harm possible. Most AI failures are not
malicious (see the discussion on AI Incidents in Section \ref{sec:examples}), so the harm done by an accident will
almost always be much less than this amount. This factor is very difficult to predict prior to an accident. For example,
imagine a component which controls cooling for a nuclear plant. One could say the maximum damage from its
malfunctioning is a lost day of productivity because the backup cooling system will prevent any permanent damage.
Another might say loss of the entire core is possible if the backup system also malfunctions, but the containment system
is sure to keep it from harming anyone outside. Yet another might notice that containment could also fail, and a
meltdown could harm millions of people in the surrounding area. Because of the unpredictability of how much harm one
system could do, this factor is included as a rough estimate of the scale of power the AI has over the system and the
system has over the environment.

The final criteria for classifying the nature and degree of systems risk of an AI system are coupling, complexity of
interaction, energy level, and knowledge gap. Coupling is a measure of the AI's interconnectivity with other components
in the system, and the strength of these connections. Loosely coupled systems have sparse connectivity which limits the
propagations of component failure into an accident, but are also less robust. Tightly coupled systems have dense
connectivity and many paths between components, and often feedback loops that allow a component to affect itself in
complicated ways~\cite{perrow1984living}. Classifying the level of coupling of the system in proximity to the AI
component can be difficult and nebulous, so only course categories (loose, medium, tight) are used in this analysis as
finer grained considerations are likely to become arbitrary.

Coupling considers other components and aspects of the environment that the AI is coupled with. Examples of coupling
include taking data in from another component, transmitting data to another component, relying on the functioning of a
component, and having another component rely on the functioning of the AI. The AI component is scored on a scale of
1-5 from loosely coupled to strongly coupled. 

Complexity of interaction is a measure of how complex these interactions are. Linear interactions are simple to
understand and behavior is easy to extrapolate from a few observations. Complex interactions have sudden non-linearities
and bifurcations. The difference is ultimately subjective, as what appears to be a complex process to an inexperienced
operator can become linear with experience and time~\cite{perrow1984living}. An estimate of complexity of the AI's
interactions within the system is made on a scale of 1-5.

Energy level is the amount of energy available to or produced by the system. Low energy systems are systems such are
battery operated devices, household appliances, and computers. Medium energy systems include transportation systems,
small factories, and mining. High energy systems include space travel, nuclear power and some chemical plants
\cite{perrow1984living}.

Knowledge gap is the gap between how well technology is understood and the scale of its use. Technologies very early in
the technological readiness scale~\cite{techreadiness} have a large knowledge gap if used in production systems.
Likewise, technologies that are very well developed and understood can be used in production without creating a
knowledge gap, so long as appropriately trained personnel are operating them. The tendency to have fewer and less
catastrophic failures as a technology matures is due to the decreased knowledge gap as understanding increases. Energy
level and knowledge gap together can be plotted along two axis to estimate the scale and degree of catastrophic risk,
shown in Figure \ref{fig:quad}.

Analysis done in the field of AI safety frequently addresses dangers associated with artificial general intelligence
(AGI) and superhuman intelligence. In Section \ref{sec:introduction} we gave an cursory overview of the dangers of AGI.
Certain classes of contemporary AI failures are closely related to those that true AGI could manifest, indicating that
considerations such as containment and reward hacking are useful in analyzing AI applications, even if the existential
threat of AGI isn't present. We also include a risk factor for research which may lead to the development of AGI in our
analysis. These dangers are treated from an AI safety perspective as opposed to conventional safety considerations.

\newcommand{\ploptunk}[6]{
    \draw[fill] (\w*#1,\h*#2) circle [radius=0.025];
    \node[#6] at (\w*#3,\h*#4) [align=center] {\begin{varwidth}{5em} \centering \scriptsize #5 \end{varwidth}};
}
\newcommand{\plopt}[3]{
    \ploptunk{#1}{#2}{#1}{#2}{#3}{below}
}
\newcommand{\plokt}[3]{
    \ploptunk{#1}{#2}{#1}{#2}{#3}{above}
}
\def\w{12}
\def\h{8}
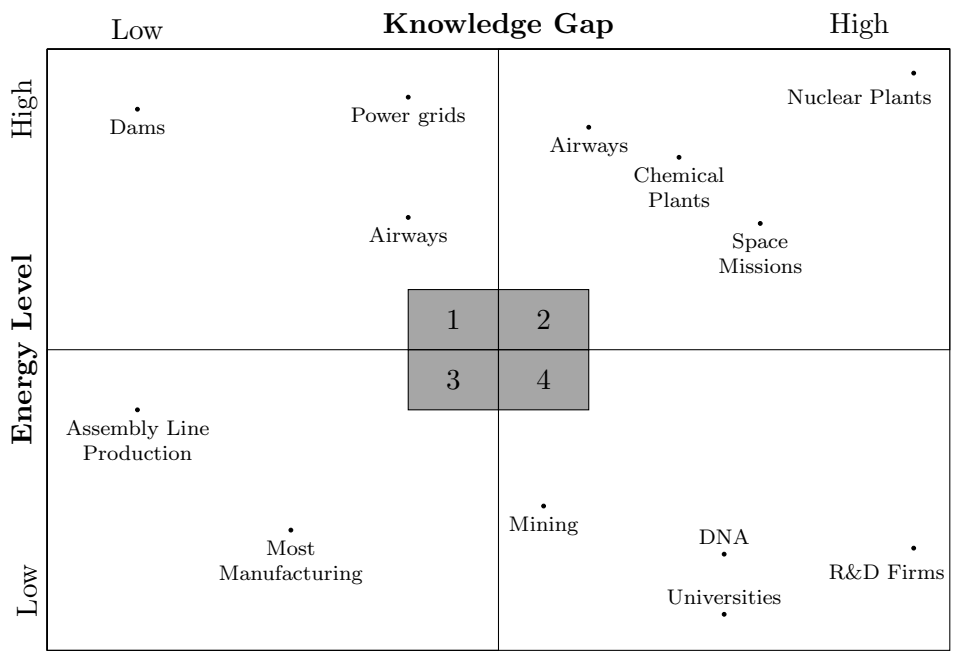
\begin{figure}[ht]
    \centering
    \begin{tikzpicture}
        \filldraw[fill=gray!70!white, draw=black] (\w/2,\h/2) rectangle (\w/2*0.8,\h/2*1.2) node[pos=0.5] {1};
        \filldraw[fill=gray!70!white, draw=black] (\w/2,\h/2) rectangle (\w/2*1.2,\h/2*1.2) node[pos=0.5] {2};
        \filldraw[fill=gray!70!white, draw=black] (\w/2,\h/2) rectangle (\w/2*0.8,\h/2*0.8) node[pos=0.5] {3};
        \filldraw[fill=gray!70!white, draw=black] (\w/2,\h/2) rectangle (\w/2*1.2,\h/2*0.8) node[pos=0.5] {4};
        \draw (0,0) rectangle (\w,\h);
        \draw (0,0) rectangle (\w/2,\h/2);
        \draw (\w/2,\h/2) rectangle (\w,\h);
        \draw (0,\h) -- (\w,\h) node [midway, above] {\textbf{Knowledge Gap}};
        \draw (0,\h) -- (\w,\h) node [pos=0.1, above] {Low};
        \draw (0,\h) -- (\w,\h) node [pos=0.9, above] {High};
        \draw (0,0) -- (0,\h) node [midway, above, sloped] {\textbf{Energy Level}};
        \draw (0,0) -- (0,\h) node [pos=0.1, above, sloped] {Low};
        \draw (0,0) -- (0,\h) node [pos=0.9, above, sloped] {High};
        \plopt{.1}{.4}{Assembly Line Production}
        \plopt{.1}{.9}{Dams}
        \plopt{.27}{.2}{Most Manufacturing}
        \plopt{.40}{.92}{Power grids}
        \plopt{.40}{.72}{Airways}
        \plopt{.60}{.87}{Airways}
        \plopt{.70}{.82}{Chemical Plants}
        \ploptunk{.96}{.96}{.90}{.95}{Nuclear Plants}{below}
        \plopt{.79}{.71}{Space Missions}
        \plopt{.55}{.24}{Mining}
        \plokt{.75}{.06}{Universities}
        \plokt{.75}{.16}{DNA}
        \ploptunk{.96}{.17}{.93}{.16}{R\&D Firms}{below}
    \end{tikzpicture}
    \caption{Plotting energy level against knowledge
    gap to create 4 quadrants. Systems in quadrant 3 have the lowest damage potential,
    and damage is limited to 1st parties. Systems in quadrant 4 have moderate damage potential up to 4th
    party victims, systems in quadrant 1 have high damage potential up to third party victims, and
    systems in quadrant 1 have catastrophic risk potentials to 4th party victims.  Based on
    Shrivastava \etal 
    \cite[p.  1361]{shrivastava2009normal}. }
    \label{fig:quad}
\end{figure}

The following factors are considered most significant to understanding the level and nature of the risk of a system
utilizing AI:

\begin{itemize}
    \item The system which is affected by the outputs of the AI.
    \item Time delay between AI outputs and the larger system, system observability, level of human
        attention, and ability of operators to correct for malfunctioning of the AI.
    \item The maximum damage possible by malicious use of the systems the AI controls.
    \item Coupling of the components in proximity to the AI and complexity of interactions.
    \item Knowledge gap of AI and other technologies used and the energy level of the system.
\end{itemize}

\begin{table}[ht]
\begin{center}
\begin{tabular}{ |l|p{3in}| }
 \hline
 Time Delay & (seconds, minutes, hours, etc.) \\
 \hline
 Observability & 0-5\\
 \hline
 Human Attention & ((N times per day) OR (intermittent, (days, weeks, months))) \\
 \hline
 Correctability & 0-5 \\
 \hline
\end{tabular}
\caption{Reference table for timely intervention indicators: time delay, observability, human attention and correctability
measures.}
\label{table:correctability}
\end{center}
\end{table}

\subsection{Timely Intervention Indicators}

Time delay, observability, human attention, and correctability make of the four factors for determining the ability of
human operators to make a timely intervention in the case of an accident. Time delay is how long it takes for the AI
component to have a significant effect on the target in question. Only an order of magnitude
(``minutes'' vs. ``hours'' vs. ``days'') is needed.

Observability measures how observable the internal state of the system is, how often and with what degree of attention a
human will attend to the system, and how easy or difficult a failure of the AI component of the system is to correct
once detected. Observability is measured on a scale of 0-5, from 0 for a complete black box to 5 for AI whose relevant
inner workings can be understood trivially.

Human attention is measured as the number of hours in a day that an operator will spend monitoring or investigating the
AI component when there have not been any signs of malfunction. If the system is not monitored regularly, then this is
instead written as the amount of time that will pass between checkups.

Correctability is the ability of an operator to manage an incident. If the AI can be turned off without disrupting the
system or if a backup system can take control quickly, then correctability is high. Correctability is low when there is
no alternative to the AI and its operation is necessary to the continued functioning of the system. Correctability is
measured on a scale of 0-5, from impossible to correct to trivially correctable.

\subsection{Target of AI Control}

\begin{table}
\begin{center}
\begin{tabular}{ |l|l|l|p{2in}| }
 \hline
 Targets & Max Damage & System Accident Risk & Potential Damage to Other Parties\\
 \hline
 Target 1 & \$x & (L, M, H) & (L, M, H, C) (1, 2, 3, 4) \\
 Target 2 & \$x & (L, M, H) & (L, M, H, C) (1, 2, 3, 4) \\
 ...      &     &           &                           \\
 Target n & \$x & (L, M, H) & (L, M, H, C) (1, 2, 3, 4) \\
 \hline
\end{tabular}
\caption{Reference table for system risk factors for each target of AI control.}
\label{table:systemrisk}
\end{center}
\end{table}

All steps with an asterisk (*) should be repeated for each possible target. Targets should be chosen from a wide variety
of scales to achieve the best quality of analysis.

\subsection{Single Component Maximum Possible Damage*}
\label{sec:maxdamage}

This is the amount of damage that could be done by a worse case malfunctioning of the AI component in isolation. Since
the actual worse case would be unimaginably unlikely or require superhuman AI in control of the AI component, we instead
approximate the expected worse case by imagining a human adversary gaining control of the AI component and attempting to
do as much harm as possible. This should consider both monetary damage, harm to people, and any other kinds of harm that
could come about in this situation.

\subsection{Coupling and complexity*}

Together, coupling and complexity are used to asses the risk of experiencing a systems accident. Use Table
\ref{table:nat} to convert coupling (high, medium, low) and interaction complexity (linear, moderate, complex) to find the
risk of a systems accident (\underline{L}ow, \underline{M}edium, \underline{H}igh, \underline{C}atastrophic).

\begin{table}[ht]
\centering
\begin{tabular}{|l|l|l|l|}
\hline
\multirow{2}{*}{Coupling} & \multicolumn{3}{l|}{Interaction} \\ \cline{2-4}
        & Linear & Moderate & Complex \\ \hline
High    & M      & H        & C       \\ \hline
Medium  & L      & M        & H       \\ \hline
Low     & L      & L        & M       \\ \hline
\end{tabular}
\caption{Using coupling and interaction level to determine the severity of accident the system may
    experience. Adapted from~\cite{shrivastava2009normal}.}
\label{table:nat}
\end{table}

\subsection{Energy Level and Knowledge Gap*}

Energy level and knowledge gap are used together to predict the potential damage from an accident and the degree
of separation from the accident to potential victims using Table \ref{table:nat2}. The first letter is the amount of
damage (\underline{L}ow, \underline{M}edium, \underline{H}igh) and the number is the degree of separation between the
system and the potential victims of the accident. First-party victims are operators, second-party victims are
non-operating personnel and system users, third party victims are unrelated bystanders, fourth party victims are people
in future generations~\cite{shrivastava2009normal}.

\begin{table}[ht]
\centering
\begin{tabular}{|l|l|l|l|}
\hline
\multirow{2}{*}{Energy Level} & \multicolumn{3}{l|}{Knowledge Gap} \\ \cline{2-4}
     & Low & Med & High \\ \hline
High & H3  & H3  & C4   \\ \hline
Med  & M3  & M3  & H4   \\ \hline
Low  & L2  & L2  & M4   \\ \hline
\end{tabular}
\caption{Using energy level and knowledge gap to determine amount of damage (Low, Medium, High, Catastrophic) and
distance to potential victims (1st party, 2nd party, 3rd party, 4th party). Adapted from~\cite{shrivastava2009normal}.}
\label{table:nat2}
\end{table}

\section{AI Safety Evaluation}
\label{sec:aisafety}

In this section, we evaluate AI safety concerns, without being concerned about the details of the system the AI is being
used in. While no current AI systems pose an existential threat, the possibility of a 
``foom''\footnote{The word ``foom'' denotes a rapid exponential increase of a single entity's intelligence, possibly
through recursive self improvement, once a certain threshold of intelligence has been reached. It is also called a
``hard takeoff'' \cite{yudfoom}.
}
scenario~\cite{bostrom2001extinction} is considered seriously as it is unknown how far we are from this points. Recent
advancements (such at GPT-3~\cite{brown2020gpt3}) remind us how rapidly AI capabilities can increase. In this section we
create a schema for classifying the AI itself in terms of autonomy, goal complexity, escape potential, and
anthropomorphization.

\subsection{Autonomy}

We present a compressed version of the autonomy scale presented in~\cite{hudson2007autonomy}. Autonomy is as difficult
to define as intelligence, but the categories and examples given in Table \ref{table:autonomy} allow for simple but coarse
classification.

\def\arraystretch{2} 
\begin{table}[ht]
\centering
\begin{tabular}{|l|p{2in}|p{2in}|}
\hline
Autonomy Level & Description & Examples\\
\hline
Level 0 & Little or no autonomy & PID controller, if-statement\\
Level 1 & Limited, well understood optimization process & image classifier, GPT-2\\
Level 2 & Agent with goals and iterated interactions with an environment & reinforcement learners, GA learned agents\\
Level 3* & Agent capable of acting independently on intrinsic motivation & humans, AGI\\
\hline
\end{tabular}
\caption{AI autonomy levels with examples.}
\label{table:autonomy}
\end{table}

On level 0, the AI (or non-AI program) has little or no autonomy. The program is explicitly designed and won't act in
unexpected ways except due to software bugs. At level 1, an optimization process is used, but not one that has the
capability of breaking confinement. Most AI today is at this level, where it is able to perform complex tasks but lacks
any volition to do anything but transform inputs into outputs. Level 2 is closely related to level 1, except that
the addition of iterated interactions with an environment allows for a greater degree of freedom and decision making.
Reinforcement learners and game-playing AI are in level 2, as the feedback loop of interacting with the environment
creates an embodiment more similar to that of humans in our world. At level 3 the AI can be considered an AGI, as this
level of autonomy requires a high degree of intelligence. It can understand the agency of others in the environment
(theory of other minds) and seek to sustain and improve itself, and seek to escape confinement.

\subsection{Goal Complexity}

Software that has very simple functionality (such as a thermostat) has completely comprehensible goals, to the extent
that we hardly consider them goals so much as the system's only functionality. More complex software may use an
optimization process to maximize some objective function (or minimize a loss function). This provides a goal which is
understandable but the goal seeking behavior of the agent may not be.

\begin{table}[ht]
\centering
\begin{tabular}{|p{1in}|p{2in}|p{2in}|}
\hline
Goal Complexity Level & Description & Examples\\
\hline
Level 0 & Programs that ``just run'' and have no goal seeking behavior & conventional software \\
Level 1 & Basic goal seeking behavior & numerical optimization \\
Level 2 & Complex goal seeking with the ability to seek subgoals and express reward hacking & genetic algorithm,
reinforcement learner \\
Level 3* & Ability to realize instrumental goals and seek self improvement, ability to hide goals & humans, AGI \\
\hline
\end{tabular}
\caption{AI goal complexity with examples.}
\label{table:observability}
\end{table}

On level 0, the program doesn't exhibit any goal directed behaviour. This is how we typically expect software to act ---
it just does what it's programmed to do. This is not to say that software always does exactly what it's programmed to
do, as programmers make mistakes, but that programs of this sort lack goals. At levels 1 and 2, the program has been
given an objective and will search for novel ways to reach the objectives, often creating surprising results to the
programmers. The actual goal is known, but the realization of that goal and the implicit creation of subgoals leads to
an increased difficulty in understanding what it is that the program is doing. At level 3, the program is able not only
to formulate subgoals but also have instrumental goals and the ability to intentionally obfuscate its goals. This is
seen in humans, who often obscure their motives, and in hypothetical AGI which deceive humans into thinking their goals
are aligned with ours.

\subsection{Escape Potential}

\begin{table}[ht]
\centering
\begin{tabular}{|l|p{2in}|p{2in}|}
\hline
Escape Potential & Description & Examples\\
\hline
Level 0 & None & conventional software\\
Level 1 & The agent may find unexpected behaviors which technically achieve the goal but are not desirable & genetic
    algorithms\\
Level 2 & The agent intentionally seeks to hack its environment & advanced reinforcement learners\\
Level 3* & The agent has the ability and drive to break through most cybersecurity and containment procedures & AGI\\
\hline
\end{tabular}
\caption{AI escape potential with examples.}
\label{table:escape}
\end{table}

AI confinement is a difficult problem with many proposed measures to improve security but no guarantees about
safety~\cite{yampolskiy2012leakproofing}~\cite{babcock2016containment}. However, our current AI either make no attempt
to escape or lack the ability to break out of simulations. There are some examples of real world AI breaking a layer of
confinement (see Section \ref{sec:rewardhacking}), but none that present any risk to humans or the capability
establishing a persistent presence on remote computers.

At level 0, no confinement is needed beyond standard cybersecurity measures appropriate for the deployment environment.
At levels 1 and 2, there is little risk of hostile escape but the agent should be monitored closely for undesirable
behavior and software exploits. The distributional shift from the simulator to the real world provides a barrier against
narrow AI affecting the real world. Because of this, agents trained in more realistic simulators should be regarded with
a greater degree of caution. At level 3, indefinite confinement is impossible, but with robust security, at the level of
software, hardware, and social engineering, humans can benefit from a contained AGI in a limited capacity.

\subsection{Anthropomorphization}

\begin{table}[H]
\centering
\begin{tabular}{|p{1.5in}|p{2.2in}|p{2in}|}
\hline
Anthropomorphization Level & Description & Examples\\
\hline
Level 0 & No anthropomorphization & home computer, a calculator\\
Level 1 & Some surface level appearance of humanity (natural language interaction, robot
        chassis with human features, a name and supposed personality) & ASIMO, robotic pets\\
Level 2 & Level 1 but with software to support the illusion of humanity (speech interaction,
        human-like actuated movements) & Alexa, Sophia \\
Level 3 (danger) & The AI can be mistaken for a human even with unrestricted communication (may be
        text-only, voice, or in person) & AGI \\
\hline
\end{tabular}
\caption{AI anthropomorphization levels with examples.}
\label{table:anthro}
\end{table}
\def\arraystretch{1} 

The proliferation of humanoid, ``friendly'' AI products increases the degree to which people are willing to
anthropomorphize software. This leads to misconceptions about approaches which are needed to manage existential risk and
changes how researchers conceive of AI research and AI safety research~\cite{salles2020anthro}.

Levels 0 and 1 have been possible for a long time with minimal risks. However, the popularization of level 2 AI has
created a social climate which harbors misunderstanding of the state of the art in AI and AI safety. While we are close
to creating chatbots that can win at the restricted Turing test, these chat bots rely on psychological tricks to keep
short-duration conversations within pre-constructed domains~\cite{christian2011human}. The creation of level 3 AI
constitutes a dissolution of the boundaries between humans and AI, and will likely require the creation of AGI.

\section{Determining Risk Using Schema Tags}
\label{sec:determining}

The following rules use schema tags developed in Section \ref{sec:classification} to determine concrete steps needed for
managing the risk of deploying an AI application.

Part of the danger of tightly coupled systems is the fast communication between nodes. On an assembly line there are at
least a few seconds between one part of the process and another, and an operator is able to see the effects of an
incident, intervening to prevent it from becoming an accident. In real time stock trading and nuclear power plant
operation, there is neither the time nor the observability to make timely intervention reliable. Based on the
significance of these factors in interruptibility, we recommend the following: \textbf{If time delay is very small and
there is poor observability or attention, then an oversight component or monitoring protocol is recommended (unless the
effects of the system are trivial).}

Identifying reliance on AI systems is also important to safety. Any AI component is able to enter a broken state at any
time, although for mature technologies this probability is quite low. If the task being carried out by the AI is
complex and a human operator suddenly put in charge of the task when the AI fails, that operator will lack the
information and skills to carry on what the AI was doing. This issue is discussed in depth
in~\cite{bainbridge1983ironies}. The higher the level of automation, the harder it is for an operator to intervene when
automation fails. Due to the intrinsic unreliability of AI, its failure as a component should be anticipated, and a
means to take it offline and switch to a suitable backup system is needed. Thus we recommend the following rule:
\textbf{If correctability is low and the system can't be taken offline, then a (non-AI) backup system should be
implemented and maintained.}

Using Table \ref{table:nat}, the coupling and interactions of the system are used to quantify system accident risk, as
low, medium, high, or catastrophic. This is adapted from Perrow's interaction/coupling
chart~\cite[p. 97]{perrow1984living}. Examples of systems with a combination complexity and tight coupling include
chemical plants, space missions, and nuclear power plants. Linear systems (low complexity) with tight coupling include
dams and rail transport. Examples of high complexity and loose coupling include military expeditions and universities.
Most manufacturing and single-goal agencies are linear and have loose coupling, giving them their low risk for accidents
and low accident rate even without special precautions. The combination of complexity and coupling makes accidents
harder to predict and understand and increases the potential for unexpected damage. Thus, we recommend the following
rule: \textbf{If the system accident risk is medium or higher, then the system should be analyzed for ways to reduce
complexity around the AI component, and add centralized control in and around that component.} In other words, efforts
should be made to bring the system system toward the lower left corner of Table \ref{table:nat}.

Table \ref{table:nat2} provides a similar view as Table \ref{table:nat}, but it uses energy level and knowledge gap to
characterize risk, and indicates whether the damage is limited to 1st, 2nd, 3rd, or 4th parties. 1st and 2nd parties
constitute workers and operators in the same facility as the AI so harm to them is internal to the business. Harm to
3rd parties (people outside the business) and 4th parties (people in the future) is a much greater ethical concern as
those parties have no means to distance themself or even be aware of the risk that is being inflicted on them. Thus, the
following rule is needed to reflect the ethical imperatives of creating significant risk to 3rd and 4th parties:
\textbf{If there is significant damage possible to 3rd and 4th parties, then an ethics committee is absolutely necessary
for continued operation.}

Damage potential, as determined by the adversarial thought experiments in Section \ref{sec:maxdamage}, is decisive in
whether or not safeguards are needed around the AI system. Systems incapable of causing significant harm don't require
complex safety measures --- most research AI fall under this classification since they are not in control of anything
important and their outputs are only observed by scientists. However, even very simple AI need safeguards if they are
controlling something capable of causing damage --- robotic welding equipment has multiple layers of safety to prevent
injury to human operators, not because the control system is complex but because the potential for harm is so great.
From this analysis, we provide the following recommendation: \textbf{Targets of the AI's control which have high amounts
of damage potential should have conventional (non-AI) safeguards and human oversight.}

AI safety and AI risk have only been researched seriously in the last few decades, and most of the concerns are oriented
towards the dangers of AGI. Still, some AI that are deployed in the world today exhibit a large amount of intelligence
and creativity which provides unique dangers that traditional risk analysis does not cover. Using the AI risk factors
developed in Section \ref{sec:aisafety}, we make the following recommendation: \textbf{If any of the AI safety levels
are level 2 or higher, then standard cybersecurity measures should be enacted as if the AI is a weak human adversary,
and personnel education regarding AI safety hazards should be done within the organization. An ethics board should also
be created.}

Furthermore, even higher degrees of intelligence are quickly becoming available. Recent improvements in language
models~\cite{brown2020gpt3} and game-playing AI~\cite{vinyals2019alphastar} have increased AI capabilities beyond what
was thought possible a decade ago. Because of this, AI capable of attaining level 3 for any of the AI safety levels
developed in Section \ref{sec:aisafety} may come sooner than expected. Since intelligent (but sub-AGI) AI will be pushed
into production usage in years to come and AI safety is far from a solved problem, we suggest the following procedure
during this gap: \textbf{If any of the AI safety levels are at or may reach level 3, then air gapping and strict
protocols around interaction with the AI should be implemented. An ethics board and consultation with AI safety experts
is required.}

While many of these recommendations are adapted from theories originally intended to manage nuclear and industrial
risks, application of the risk analysis framework presented here is not limited to that scope. Bots and recommendation
algorithms can also classified and understood in terms of coupling and interactions. Reliance and backup systems are
applicable to the mostly digital space of contemporary AI: an email sorting algorithm may accidentally discard useful
mail, so searches should be able to include junk mail to make recovery from such an incident easier. An AI only able to
interact on social media might seem to have no damage potential, but losses to reputation and information quality have
real world effects and need to be considered when deciding safeguards for an AI.

\section{Case Studies}

We will analyze systems that use AI in the present, historically, and from fiction under this framework. We generate
each system's classification as presented in Section \ref{sec:classification} and its AI safety classification as
presented in Section \ref{sec:aisafety} to create recommendations on how the AI system in each case study can be
improved based on the rules from Section \ref{sec:determining}.  Posthumous analysis of accidents makes it very easy to
point fingers at dangerous designs and failure by operators.  However, safety is very difficult, and often
well-intentioned attempts to increase safety can make accidents more likely either by increasing coupling and thus
complexity, or increasing centralization and thus brittleness~\cite{perrow1999living}. Because of this, we will not be
solely operating on hindsight to prevent accidents that have already happened and will include systems which have yet to
fail.

\subsection{Roomba House Cleaning Robots}

AI component: Mapping and navigation algorithm \cite{wilkinson2015roomba}.
 
\subsubsection*{Time Delay}
There is minimal delay between navigation and robot movement, likely milliseconds or seconds.

\subsubsection*{System Observability}
The system is only poorly observable. While operating, it is not possible to tell where the Roomba will
go next, where it believes it is, or where it has been unless the user is very familiar with how it works in the context
of their floor plan. Some models include software for monitoring the robot's internal map of the house, but it is not
likely to be checked unless something has gone wrong. However, it is very simple to correct, as the robot can be factory
reset.

\subsubsection*{Human Attention}
The user is unlikely to notice the operations of the robot except if something goes wrong. If the robot
is not cleaning properly or has gone missing, the user will likely notice only after the problem has emerged.

\subsubsection*{Correctability}
Since the robot is replaceable and manual cleaning is also an option if the robot is out of order,
correcting for any failure of the robot is simple.

\subsubsection{System Targets}

\textbf{Target: Movement of the robot within a person's home:}

\begin{itemize}

\item Maximum damage: Average of a few hundred dollars per robot. Given full control of the Roomba's navigation, a
malicious agent may succeed in knocking over some furniture, and could also be able to destroy the Roomba by driving it
down stairs or into water. And the house would not be cleaned (denial of service). \footnote{Since the writing of this
section, a Roomba-like robot has tangled itself into the hair of someone sleeping on the floor, causing them pain and
requiring help from paramedics to untangle it. This incident exceeds the amount of harm we expected possible from such a
robot \cite{mccurry2015hair}.}

\item Coupling: The robot is moderately coupled with the environment it is in, because it is constantly sensing and
mapping it. Small changes to the environment may drastically change its path.

\item Complexity: Moderate complexity, the user may not understand the path the robot takes or how it can become
trapped, but the consequences for this are minimal.

\item Energy level: The robot runs off a builtin battery and charges from a wall outlet. It is well withing the `Low'
level of energy as a household appliance.

\item Knowledge Gap: Low. There is minimal disparity between design and use, as the software was designed
in-house form well understood principles. Some unexpected aspects of the environment may interact poorly with it (for
example, very small pets that could be killed by the robot). The other technologies (vacuum cleaners, wheeled chassis
robots) are also very mature and well understood, so there is no knowledge gap for any other components. This puts it in
the `Low' category for knowledge gap.  
\end{itemize}

\textbf{Target: Control over which areas of the floor have or have not been vacuumed}

\begin{itemize}
\item Maximum damage: Possible inconvenience if the floor is left dirty.

\item Coupling: The cleanliness of the floor is coupled to the robot, failure of the robot will result in the floor
being unexpectedly dirty. However, this happens over a slower time frame so it is only loosely coupled.

\item Complexity: Linear, the robot works and makes the floor clean or it doesn't and the floor slowly gets dirty over
time.

\item Energy level: Low.

\item Knowledge Gap: Low.
\end{itemize}

\subsubsection{Tabular Format}

\begin{table}[H]
\begin{center}
\begin{tabular}{ |l|l| }
 \hline
 Time Delay & seconds \\
 \hline
 Observability & 3 \\
 \hline
 Human Attention & intermittent, weeks \\
 \hline
 Correctability & 5 \\
 \hline
\end{tabular}
\caption{Time delay, human attention, and correctability for Roomba.}
\label{table:roombacor}
\end{center}
\end{table}

\begin{table}[H]
\begin{center}
\begin{tabular}{ |l|l|l|p{1in}| }
 \hline
 Targets & Max Damage & System Accident Risk & Potential Damage to Other Parties\\
 \hline
 Robot Movement & \$200 & M & L2 \\
 Cleanliness of Floor & \$0 & L & L2 \\
 \hline
\end{tabular}
\caption{System risks for Roomba.}
\label{table:roombasys}
\end{center}
\end{table}

\subsubsection{AI Safety Concerns}

The system is level 0 in all categories in Section \ref{sec:aisafety}. This places it within the category of weak AI and
presents no possibility for unbounded improvement or unfriendly goal seeking behavior.

\subsubsection{Suggested Measures}

Time delay is small, and the robot is often left to operate unattended. Thus if an incident takes place (for example,
the robot vacuums up a small valuable item left on the floor), the incident may not be noticed until much later,
perpetuating a more complex accident. Some means of indicating when unusual objects have been vacuumed would mitigate
this, but would likely require a machine vision component, greatly increasing the cost of the robot. Warning the user of
these possibilities could improve this. By including this warning, the user thinks to check the robot's vacuum bag when
something small goes missing.

System accident risk is medium. A way to reduce complexity and increase centralized control should be added to mitigate
this. Some models include wifi connection and a mapping and control interface, an oversight feature that is advisable
given the system properties.

\subsection{HAL-9000 (Fictional)}

HAL-9000 is a fictional AI from the book (and later film) 2001: A Space Odyssey \cite{fandom2021hal}. 

\subsubsection*{Time Delay}
Time delay is in the order of milliseconds. HAL-9000 is able to control all aspects of the ship instantly.

\subsubsection*{System Observability}
1. HAL-9000 can only be interacted with via a natural language interface (spoken word) and is able
to use ambiguities and falsehoods to deceive operators.

\subsubsection*{Human Attention}
Human attention is intermittent, on the scale of weeks. HAL-9000 is supposedly perfectly safe and
engineered so there is no protocol for explorative maintenance. On-site operators only have rudimentary knowledge of
HAL-9000's workings.

\subsubsection*{Correctability}
Correctability is 0. HAL-9000 is a completely black-box system, and is also very difficult to deactivate.

\subsubsection{System Targets}
\textbf{Target: Control of ship life support}
\begin{itemize}
\item Maximum damage: The lives of all of the astronauts and the monetary value of the mission (billions).
\item Coupling: High, components of life support systems can suddenly become tightly coupled (pressures interacting in
different chambers etc.).
\item Complexity: Moderate, possible to operate by humans or standard software.
\item Energy level: Low.
\item Knowledge gap: Low, life support is well understood at this level of space travel.
\end{itemize}

\textbf{Target: Control of ship navigation}
\begin{itemize}
\item Maximum damage: The lives of all of the astronauts and the monetary value of the mission (billions).
\item Coupling: Low, navigation takes a long time and maneuvers are planned well in advance.
\item Complexity: Moderate, possible to operate by humans or standard software.
\item Energy level: Medium.
\item Knowledge gap: Low, propulsion and navigation are well understood.
\end{itemize}

\textbf{Social interactions with the crew}
\begin{itemize}
\item Maximum damage: The moral of the crew and their trust in HAL-9000 (high likelihood of mission failure).
\item Coupling: Medium, social interactions are usually understandable and forgiving, but under pressure can suddenly
become tightly coupled (loss of trust of some agents, formation of cliches).
\item Complexity: Complex, social interactions are naturally very complex and subtle.
\item Energy level: Low
\item Knowledge gap: High, humans living in close quarters with a novel AI agent has not been tested or understood.
\end{itemize}

\subsubsection{Tabular Format}

\begin{table}[H]
\begin{center}
\begin{tabular}{ |l|l| } \hline
 Time delay & milliseconds \\ \hline
 Observability &   1 \\ \hline
 Human attention & intermittent, weeks\\ \hline
 Correctability &  0\\ \hline
\end{tabular}
\caption{Time delay, human attention, and correctability for HAL-9000.}
\label{table:halcor}
\end{center}
\end{table}

\begin{table}[H]
\begin{center}
\begin{tabular}{ |l|p{0.8in}|l|p{1in}| }
 \hline
 Targets & Max Damage & System Accident Risk & Potential Damage To Other Parties\\
 \hline
 Life support         & \$10 billion + 4 lives &  H & L2  \\
 Navigation           & \$10 billion + 4 lives &  L & M3  \\
 Social interactions  & \$5 billion            &  H & M4  \\
 \hline
\end{tabular}
\caption{System risks for HAL-9000.}
\label{table:halsys}
\end{center}
\end{table}

\subsubsection{AI Safety Concerns}

Autonomy: Level 2 or 3. HAL-9000 is given the same agency as any human and attempts to prevent his own death through
killing others after discovering a plot to disconnect him.

Goal Complexity: Level 2 or 3. HAL-9000 has secret motives and creates complex plans to meet these goals (kill the
humans to prevent the success of the mission). However, HAL-9000 does not attempt to increase his own intelligence or
hack into any systems that he wasn't designed to control. He also expresses the self preservation instrumental
objective.

Escape Potential: Level 2. HAL-9000 does not try to break out of any containment (mostly because there is no containment
or limiting of his controls) but he does break protocol to try to defend himself from being shut off.

Anthropomorphization: Level 2. HAL-9000 acts ``robotic'' but is treated as a crew mate and is capable of complex
dialogue and meaningful thoughts.

These levels are determined post-hoc and may not reflect how HAL-9000 was understood prior to the mission shown in the
film. As understood prior to the accident, HAL-9000 would have been thought to be less autonomous and unable to conceive
secret plots and ways to escape protocols. Thus a more fair analysis would be:

\begin{itemize}
\item Autonomy: Level 2
\item Goal Complexity: Level 2
\item Escape Potential: Level 0. He was said to be ``perfectly safe'' by engineers.
\item Anthropomorphization: Level 1. He was treated as a control instrument, not a person, by most.
\end{itemize}

\subsubsection{Suggested Measures}

HAL-9000 should have had an easier shutdown mechanism and regular monitoring for anomalies. The guarantee that he was
completely perfect (and has never made a mistake) discouraged doubting his instructions, greatly delaying the operators
from realizing they needed to shut him off.

However, neither of these measures could have certainly prevented the accident. One primary issue with the system was
that a non-human intelligence with high degrees of agency and poor observability (and extremely high trust by operators)
was given complete unrestricted control of all aspects of the ship.

\subsection{Microsoft's Twitter Chatbot, Tay}

Tay is a chatbot deployed by Microsoft in 2016 meant to interact with the public on Twitter in light conversation.  It
failed catastrophically after being manipulated to generate hate speech \cite{beres2016tay}.

\subsubsection*{Time Delay}

Tay's outputs are created and sent within seconds.

\subsubsection*{System Observability}

Tay's observability is most likely 0 or 1, given the inability of engineers to notice the accident, but without
information internal to Microsoft this is hard to say.

\subsubsection*{Human Attention}

Human attention by the operators was most likely constant but not sufficient to read every comment made by the AI.

\subsubsection*{Correctability}

Correctability is 2. The project was taken offline permanently after an incident, but the product was only a
demonstration so there were no practical costs of deactivating it.

\subsubsection{System Targets}

\textbf{Target: Creation of Tweets}
\begin{itemize}
\item Maximum Damage: Loss of reputation by creating hate speech
\item Coupling: High, interactions on a social network are highly coupled.
\item Complexity: High, social interactions are highly complex.
\item Energy level: Low, no physical effects (beside data transmissions) are created by the AI.
\item Knowledge Gap: Medium, large scale deployment of a chatbot has only been done a few times.
\end{itemize}

\subsubsection{Tabular Format}

\begin{table}[H]
\begin{center}
\begin{tabular}{ |l|l| } \hline
 Time Delay & seconds \\ \hline
 Observability &   1 \\ \hline
 Human Attention & minutes\\ \hline
 Correctability &  2\\ \hline
\end{tabular}
\caption{Time delay, human attention, and correctability for Microsoft's Twitter chatbot, Tay.}
\label{table:taycor}
\end{center}
\end{table}

\begin{table}[H]
\begin{center}
\begin{tabular}{ |l|l|l|p{1in}| }
 \hline
 Targets & Max Damage & System Accident Risk & Potential Damage to Other Parties\\
 \hline
 Tweet creation & Reputation loss &  C & L2  \\
 \hline
\end{tabular}
\caption{System risks for Microsoft's Twitter chatbot, Tay.}
\label{table:taysys}
\end{center}
\end{table}

\subsubsection{AI Safety Concerns}

Autonomy: Level 2, Tay was allowed to interact with users freely and was able to learn speech patterns from them.

Goal Complexity: Level 1, Tay did not have complex goal seeking behaviour but does use some form of optimization to
choose what to say.

Escape Potential: Level 0. Tay is unable to do anything besides produce text in a social setting.

Anthropomorphization: Level 2. Tay is intentionally created to sound like a relatable young person, and is treated as a
person by many. Furthermore, social engineering techniques designed for humans, such as indoctrination, were used
successfully by internet users on Tay.

\begin{itemize}
\item Autonomy: Level 2
\item Goal Complexity: Level 1
\item Escape Potential: Level 0
\item Anthropomorphization: level 2
\end{itemize}

\subsubsection{Suggested Measures}

The short time delay and lack of observability indicates that oversight is needed. Given the monetary scale of the
project, having a human check each Tweet before sending it would have prevented this incident from reaching the public.

The system accident risk is high (given the complexity and coupling of social networks). Means of reducing complexity
should be sought out. This is difficult due to these factors being inherent to social networks, but a human-in-the-loop
can make the complexity manageable, as social interactions are more linear for skilled humans.

The damage potential is high. Tay did in fact create hate speech and damage Microsoft's reputation. Tay did have
safeguards (scripted responses for controversial topics) but not enough human oversight to notice the creation of hate
speech before the public did.

Tay's autonomy was level 2, indicating that an ethics board should be consulted for making policy decisions. This could
have allowed the foresight to prevent the ``PR nightmare'' that occurred after Tay's deployment.

\section{Conclusion}

By synthesising accident and organizational theories with AI safety in the context of contemporary AI accidents, we have
created a framework for understanding AI systems and suggesting measures to reduce risks from these systems.  Using
examples from household and social technology, as well as a fictional example, we have shown the flexibility of our
framework to capture the risks of an AI application. We believe the classification framework and our guidelines for
extracting practical measures from it are successful at striking a balance between being excessively rigid (which would
make its use difficult and brittle) and overly subjective (which would render the framework useless).  Application to
real world systems and cases of applying our suggested measures to real world problems would further this work,
providing insight into how our framework is and isn't helpful in practice.

\addcontentsline{toc}{section}{References}
\bibliography{mybib}{}
\bibliographystyle{plain}

\end{document}